\documentclass[a4paper,11pt]{article}
\usepackage{amsmath}
\usepackage{fleqn}
\usepackage{dcolumn}
\usepackage{graphicx}
\usepackage{pslatex}

\newcommand{\dt}{\delta t}
\newcommand{\Dt}{\Delta t}
\newcommand{\dtr}{\delta t_r}
\newcommand{\dts}{\delta t_\sigma}

\newcommand{\sigmaEff}{\sigma_{\mbox{\scriptsize eff}}}

\newcommand{\wInfty}{w_\infty}
\newcommand{\sigmaReal}{\sigma_{\mbox{\scriptsize real}}}

\newcommand{\Fcst}[1]{{\cal F}[ #1 ]}
\newcommand{\AvgFcst}[1]{\overline{{\cal F}}[ #1 ]}

\newcommand{\condExpt}[1]{ E\left[#1 ~|~ \Omega(t)\right]}

\newcommand{\MA}{\text{MA}}
\newcommand{\EMA}{\text{EMA}}

\begin{document} 

\begin{center}
{\LARGE\bf
       Volatility conditional on price trends
}

\vspace*{6ex}
{\Large Gilles Zumbach
    \footnote{\parbox[t]{\textwidth}{
    Consulting in Financial Research\\
    Chemin Charles Baudouin 8 \\
    1228 Saconnex d'Arve \\
    Switzerland \\
    e-mail: gilles.zumbach@bluewin.ch

    }}
}

\vspace*{4ex}
December, 2004

\vspace*{10ex}

{\Large\bf Abstract}
\end{center}
The influence of the past price behaviour on the realized volatility 
is investigated in the present article. 
The results show that trending (drifting) prices lead to
increased (decreased) realized volatility.
This ``volatility induced by trend'' constitutes a new stylized fact.
The past price behaviour is measured by a product of 2 non overlapping returns,
of the form $r\cdot L[r]$ where $L$ is the lag operator.
The effect is studied empirically using USD/CHF foreign exchange data, 
in a large range of time horizons.
A set of ARCH based processes are modified in order to include the trend effect,
and their forecasting performances are compared.
For a better forecast, it is shown that the main factor  
is the shape of the memory kernel (i.e. power law),
and the following factor is the inclusion of the trend effect.

\vspace*{8ex}
\newpage


\section{Introduction}
The extensive study of the volatility of financial time series starts 20 years ago 
with the seminal paper of Engle \cite{Engle.1982}.
After a rapid improvement with the GARCH(1,1) process \cite{RFE.1986-01-01,TBO.1986-01-01},
the quantitative results obtained since then have not been much better, 
despite the very large number of studies with various volatility processes.
Even though we have today a much better understanding of the financial markets volatility,
particularly in the high-frequency domain 
(see e.g. in \cite{IntroHighFreqFin} and references therein),
it remains difficult to translate this knowledge into better processes 
or better volatility forecasts.

In order to overcome the apparent limitation of the classical processes,
Zumbach, Pictet and Masutti \cite{GP} launched a study using genetic programming (GP).
Their work focused on improving the efficiency of the GP in order to turn it in to 
a practical tool to investigate financial time series, with an application to volatility forecast.
One important advantage of the GP is that it is not biased 
by our {\it a priori} bais and knowledge,
as the program searches in the whole space of models (yet not very efficiently).
Indeed, in \cite{GP} the GP very quickly rediscovered in essence the GARCH(1,1) model,
and then, with more time, was able to obtain better solutions.
The analysis of the solutions discovered by the 
program showed that the new terms leading to the improvement
werw of the form of a product of returns at two different time horizons.
This can be expressed in a sum of 2 terms, one with a return square,
as in most volatility processes,
and one with two non overlapping returns.
This term is like $r[\dtr](t)\cdot r[\dtr'](t-\dtr)$, 
namely at time $t$, a product of a return at time horizon $\dtr$ with a return at 
another time horizon $\dtr'$, lagged by $\dtr$ so as to not overlap the first return.
In short, we denote generically such terms as $r\cdot L[r]$, 
where $L[r]$ denotes the lagged return.
Notice that such terms are even in the return, 
namely under the change $r \rightarrow -r$, 
the term $r\cdot L[r]$ does not change its sign.

This new term can be interpreted as a measure of the past price moves, 
namely whether the market is trending or drifting.
The action of the ``trend term'' $r\cdot L[r]$ can be understood as follows.
If both returns have the same sign (both positive or both negative), 
the market is trending, 
and this may induce the market participants to change their positions because of the price move.
The trading of their positions increases volatility in the subsequent period. 
If the market is drifting, the two returns have different signs,
and the unchanged price makes the market participants to keep their positions.
This decreases the volatility in the subsequent period.
This behaviour of the market participants creates a positive correlation 
with the realized volatility in the following period.
Indeed, the correlation of the $r\cdot L[r]$ term with the realized volatility, 
namely with the volatility computed after $t$, is positive (see below).
This ``volatility induced by trend'' is a new stylized fact for financial time series. 
Notice that if a return differs by the sign but has the same magnitude,
its contribution to the historical volatility is identical.
Therefore, the trend term is measuring the recent price behaviour,
and not the historical volatility.

The goal of this paper is twofolds.
First, we would like to investigate the trend effect in empirical data.
The main questions are its magnitude, and the time horizons
of the return, lagged return and realized volatility where the trend effect is important.
Second, we want to include the effect in ARCH like processes.
The addition of such a term in a process is easy, 
it is enough to add one or several $r\cdot L[r]$ terms in the equations.
In this context, the point is not to create yet another ARCH like process, 
but to investigate the respective importance of the many ingredients 
that can enter into a volatility process.
For example, what are the respective importance of the shape of 
the memory kernel (rectangular, exponential, power law), 
the short term mean reversion, the trend effect, 
or of the various classes of market participants.
The point here is to measure the importance of 
these different stylized facts.

The paper is organized essentially along the line of the previous paragraph.
First, we describe the empirical data, 
and compute the correlation between the $r\cdot L[r]$ term and the realized volatility.
In section~\ref{section:processes}, we extend several processes with trend term(s),
while respecting the basic idea of the process (one or several components, power law, etc ...).
The processes are compared in sec.~\ref{section:volForecast} with respect to 
their forecasting performances, in light of the properties of the processes.

In many respects, this contribution is an extension of some results 
presented in \cite{GOZ.2003-10-24}. 
In particular, it will follow the same notation and processes definitions.
This paper is self-contained, but in order no to repeat extensively this reference, 
some sections are reduced to the essential.
More details can be found in \cite{GOZ.2003-10-24}, for example a discussion of the 
relative merit of the log-likelihood or of the volatility forecast, 
in relation to a given quadratic process. 

\section{The data set}
The data set used for this article is derived from 
high frequency tick-by-tick data for the foreign exchange USD/CHF.
Essentially, for the empirical analysis, 
the data set is a regular time series for USD/CHF, sampled
every 3 minutes in business time. 
First, the high frequency data is filtered for the incoherent 
effect \cite{FCO.2001-07-01}:
a very short exponential moving average is taken 
on the prices in order to attenuate the tick-by-tick 
incoherent price formation noise.
Second, the price is sampled every 3 minutes in business time.
More precisely, we use the
dynamic business time scale as developed in \cite{WAB.2000-01-31}.
Similarly to the familiar daily business time scale, the dynamic business time scale
contracts periods of low activity (night, week-end) and expands periods of high activity.
The activity pattern during the week is related to the measured volatility, 
averaged on a moving sample of 6 months. 
Holidays and day light saving time are taken into account.
The homogeneous time series used for the empirical analysis is computed from
the high frequency filtered tick-by-tick USD/CHF data, that is sampled 
using a linear interpolation, every 3 minutes in dynamic time scale.
As this market is open essentially 24 hours per day, 5 days per week,
the sampling time interval corresponds to an 
average of 2'8'' (= 5/7 3') during the 
market opening hours.  
The author is grateful to Olsen \& Associates, in Z\"urich, Switzerland,
to provide the data.

\section{Empirical analysis}
\label{sec:empiricalAnalysis}
The historical returns $r[\dtr]$ and lagged 
historical returns $L[\dtr; r[\dtr']]$ are 
computed by simple price difference from the sampled logarithmic prices.
The realized volatility is computed with 
\begin{equation}
    \sigma^2[\dts, \dtr = \dts/32] 
    	= \frac{1}{n}\sum_{t+\dtr \leq t' \leq t + \dts} r^2[\dtr](t')
	\label{eq:volDef}
\end{equation}
and where $n$ is the number of terms in the sum.
The realized volatility measures the fluctuation of the prices after $t$, 
in the interval from $t$ to $t+\dts$, 
The time horizon $\dts$, over which the volatility is computed, is the main 
volatility parameter.
The returns are taken at time horizons 
$\dtr = \dts/32$, namely they grow with $\dts$.
Other choices can be made, with a minor influence on the empirical results. 
The sum in eq.~\ref{eq:volDef} is computed with all points on the 3' sampling grid.

With the three time series $r$, $L[r]$ and $\sigma$, 
we compute the correlation
\begin{equation}
    \rho[\dtr, \dtr', \dts] = \rho(r[\dtr]\cdot L[\dtr; r[\dtr']], \sigma[\dts])
\end{equation}
where $\rho(x,y)$ on the right hand side 
denotes the usual linear correlation
between the time series $x$ and $y$.

\begin{figure}[htb]
  \centering
  \includegraphics[width=0.95\textwidth]{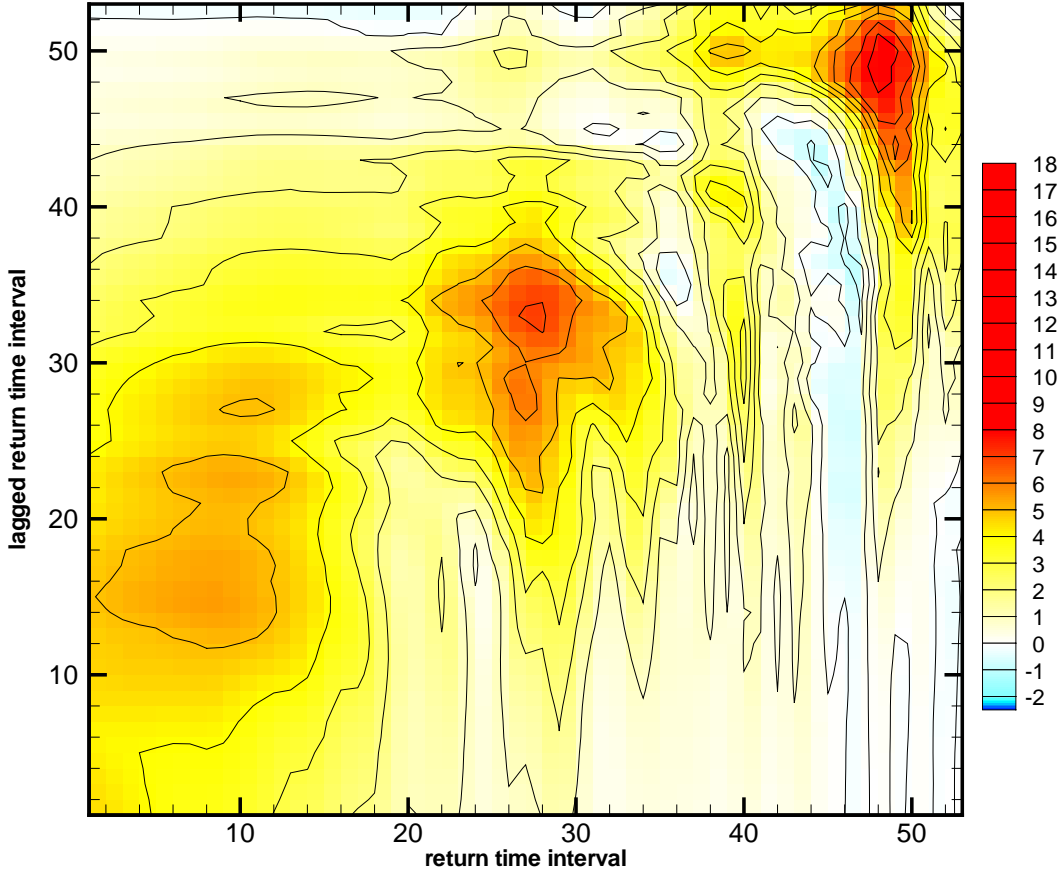}
  \caption{\small \sf Correlation between the trend term $r\cdot L[r]$ 
  and the realized volatility.
  The fixed volatility time horizon is $\dts = $1h 36. 
  The horizontal axis is the (historical) return time interval $\dtr$, 
  the vertical axis is the (historical) lagged return time interval $\dtr'$. 
The axis divisions correspond to the logarithmic of the time intervals. 
The main average physical time intervals corresponding to the labels are  
1 hour ($n=12$), 8 hour ($n=24$), 1 day ($n=31$), 1 week ($n=40$), 
and 1 month ($n=47$). Data courtesy of Olsen and Associates, Zurich.
  }
  \label{fig:fixedDts_1h36}
\end{figure}
The analysis of the empirical correlation is 
difficult to visualize, as it is a function of three time intervals. 
The 2 dimensional results presented below are cut in this 3 dimensional space,
showing the level of correlation with colors.
The figure~\ref{fig:fixedDts_1h36} is a cut at fixed volatility time interval,
at the shortest volatility estimate $\dts = 32\cdot 3' = 96'$ 
(in business time). 
The two axes correspond respectively to the return time interval 
and lagged return time interval.
The main structures emerging from this figure are:
\begin{itemize}
\item 
The correlation is essentially positive. 
This is consistent with the intuitive explanation given in the introduction,
as trends will make the market participants to modify their positions,
increasing the subsequent volatility. 
Moreover, the level of correlation is ranging from 3 to 8\%, 
corresponding to an effect of medium importance. 

\item 
We observe 4 pockets with higher correlations, 
roughly along the diagonal or above.
They are located approximately at positions 10 to 20, 30, 40 and 50,
corresponding respectively to time intervals 
intra-day, 1 day, 1 week and 1 month.
These time intervals are indeed precisely the 
expected time horizons for the main groups of traders,
in agreement with the finding of \cite{LZ.2003}.
The maxima at 1 month is very well defined, 
and with $\dtr = \dtr'$.
On the other corner, the intra-day maxima is fairly soft, 
and clearly above the diagonal.
The maxima at 1 day shows similar characteristics, whereas
the maxima at 1 week is along the diagonal and weaker. 

\end{itemize}

\begin{figure}[htb]
  \centering
  \includegraphics[width=0.95\textwidth]{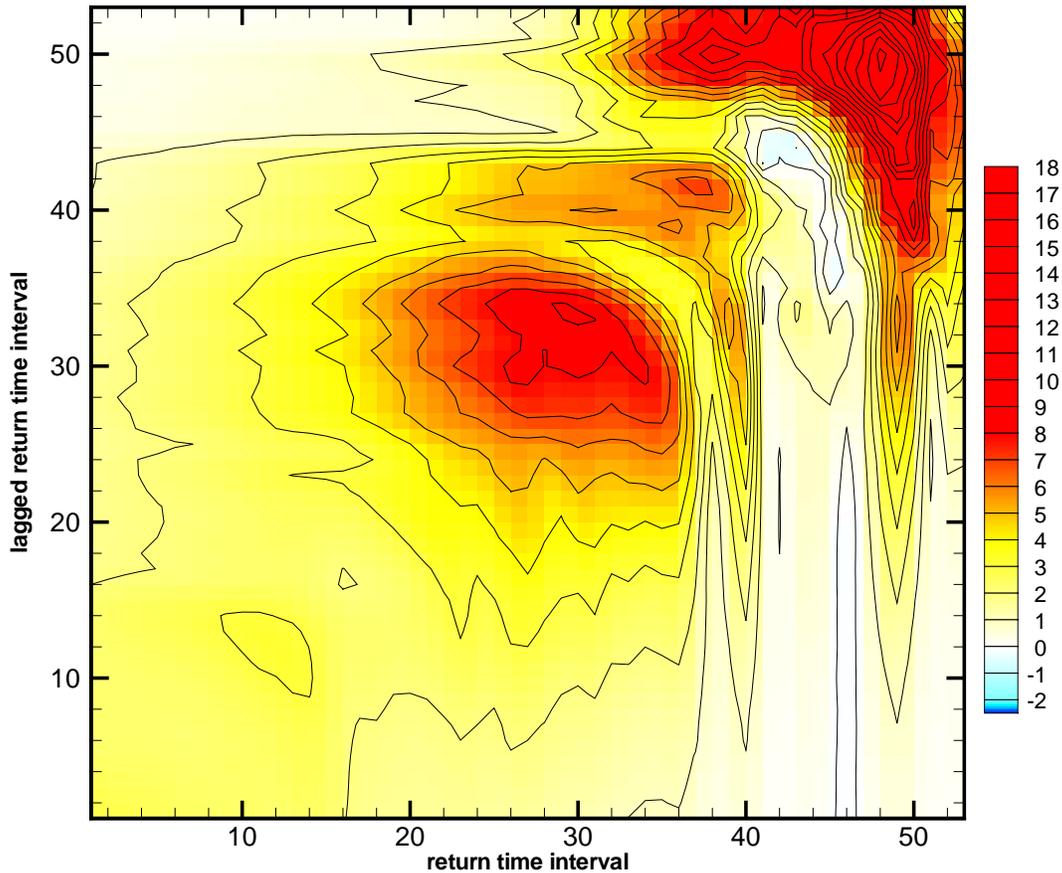}
  \caption{\small \sf Correlation between the trend term $r\cdot L[r]$ 
  and the realized volatility.
  The fixed volatility time horizon is $\dts = $1 day. 
  The axes and colors coding are as for fig.~\ref{fig:fixedDts_1h36}. 
  }
  \label{fig:fixedDts_1day}
\end{figure}
The figure~\ref{fig:fixedDts_1day} is a cut at fixed volatility time interval, 
with $\dts = $ 1 day. 
For this daily volatility, the 4 maxima can be seen, 
essentially along the diagonal $\dtr = \dtr'$. 
The intra-day maximum is very weak, the daily and monthly maxima 
are very clear, while the weekly maximum is weaker. 
This shows that the volatility at the daily time horizon 
is not influenced by intra-day trends,
but by trends at daily horizon or longer.
Both previous figures indicate that the major correlation is along the diagonal $\dtr = \dtr'$, 
or slightly above. 

\begin{figure}[htb]
  \centering
  \includegraphics[width=0.95\textwidth]{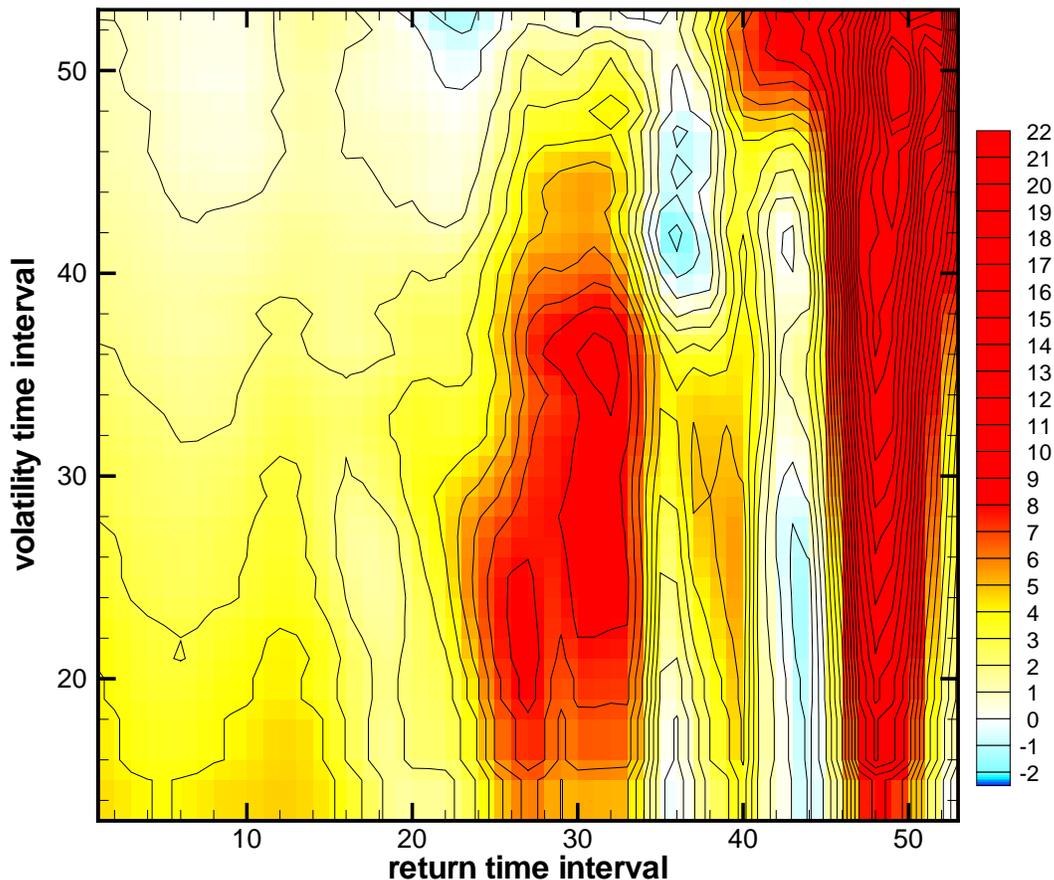}
  \caption{\small \sf Correlation between the trend term $r\cdot L[r]$ 
  and the realized volatility.
  The horizontal axis is the time horizon for both historical returns
  $\dtr = \dtr'$.
  The volatility time horizon $\dts$ is given on the vertical axis. 
  The axes divisions and colors coding are as for fig.~\ref{fig:fixedDts_1h36}. 
  }
  \label{fig:dtrEqualDtrp}
\end{figure}
The figure~\ref{fig:dtrEqualDtrp} is a cut along the plane $\dtr = \dtr'$, 
and with the volatility time horizon $\dts$ on the vertical axis.
The labels on the vertical volatility axis $\dts$ correspond to the same time intervals 
on the horizontal axis $\dtr$.
The maximum at the daily and monthly time horizons are very clear,
and the weaker intra-day and weekly maximum can also be seen.
This figure shows that trends at a given time horizon influence the volatility
up to a horizon immediately larger, and then decline.
For example, a trend measured by consecutive daily return influences strongly the 
volatility up to 3 to 4 days, but much less at one week and above.
Indeed, this behaviour is what can be expected from a portfolio manager
working on a daily horizon and adjusting its portfolio based on the 
price trends or drifts of the last few days. 

\section{Modelisation in ARCH-like processes: the ARTCH family}
\label{section:processes}
The influence of trends on the subsequent volatility 
is fairly easy to incorporate in ARCH-like processes.
Essentially, we must add one or several terms of the form $r\cdot L[r]$ 
in the conditional volatility. 
For a given ARCH-like model, there is in general one corresponding 
minimal model incorporating the influence of trends.
We call generically this new family of models by 'ARTCH', for
Auto Regressive Trend Conditional Heteroskedastic.
Let us emphasize that our goal lays not in writing new ARCH processes, 
but in quantifying various effects that can be included in a model.
To address such questions, we build a net of 
processes of increasing complexity.
All the processes are then estimated by minimizing the 1 day 
volatility forecast error on the USD/CHF data.
The comparison of their optimal forecasting performances measures 
their relative adequation to the data,
and the importance of the various effects (at the chosen time horizon). 

Our general strategy is to use quadratic ARCH-like processes 
to generate volatility forecasts.
For quadratic processes, conditional averages can be computed analytically,
and one obtains volatility forecasts that depend on the process parameters.
The properties of the forecast derive directly from the 
properties and parameters of the underlying process.
This approach gives a simple framework in which data generating processes
and volatility forecasts are closely related.
For forecasting purposes only, other direct approaches can be pursued.
Poon and Granger \cite{Poon.2003} wrote an extensive review
on volatility forecast.

Albeit the above strategy is appealing, 
the difficulty of its practical implementation should not be underestimated.
For each process, the needed conditional averages must be computed, 
and implemented numerically.
Moreover, as we are estimating the parameters by minimizing the
forecast error, the derivatives with respect to the parameters
should also be implemented. 
The theoretical setting is following closely \cite{GOZ.2003-10-24},
and only the salient points are given here for completeness.
First, we set the basic common equations, 
and then describe the various ARTCH processes.

\subsection{The basic structure of the processes}
We are considering processes for the price with the following structure
\begin{eqnarray}
	x(t+\dt) & = & x(t) + r(t+\dt)        \label{eq:base}\\
	r(t+\dt) & = & \sigmaEff(t+\dt) \epsilon(t+\dt)  \label{eq:rBase}\\
	\sigmaEff^2(t+\dt) & = & \tilde{\sigma}^2[\Omega(t), \vartheta, \dt].  \label{eq:sigmaEffBase}
\end{eqnarray}
The random variables $\epsilon$ are i.i.d. with $E[\epsilon(t)] = 0$ and $E[\epsilon^2(t)]] = 1$.
The time indexes are chosen to emphasize that $\sigmaEff^2(t+\dt)$ is a
forecast for the volatility at time $t+\dt$. 
The forecast function $\tilde{\sigma}$
is based on the information set $\Omega(t)$ at time $t$,
and depends on a set of parameters $\vartheta$.

In the processes below, the right hand side in eq.~\ref{eq:sigmaEffBase} 
contains terms of the form $r\cdot L[r]$, which have no definite sign.
As a consequence, the volatility square could become negative.
In practice, this never occurs at the optimal values of the parameters,
but it could happen during the parameters estimation.
To prevent a square root of a negative value, the right hand side 
includes a lower value threshold.
Implicit in all the ARTCH equations, a max function
$\max(\sigmaEff^2, \sigma_{min}^2)$ is included, 
where $\sigmaEff^2$ is given in the equations below. 
The minimal value for the volatility square $\sigma_{min}^2$ has value 
$\sigma_{min}^2 = 10^{-10}$.

This possible negative variance and the related $\max$ function are likely
to be an artifact of the present ARCH setting, which includes only price and volatility.
A more complete framework should include also the tick rate.
It is likely that the ``volatility induced by trend'' stylized fact is related to 
trading decisions that influence the tick rate and/or the new orders rate, 
which in turn influences the volatility.

\subsection{The GARTCH(1,1) process}
\label{sec:GARCH}
The GARTCH(1,1) process equations are
\begin{eqnarray}
    \sigma_1^2(t) & = & \mu\;\sigma_1^2(t-\dt) + (1 - \mu)r^2(t) \label{eq:sigma1-GARCH}  \\
    \sigmaEff^2(t+\dt) & = & \sigma^2 + (1 - \wInfty) \left(\sigma_1^2(t) - \sigma^2\right)
    			+ \theta ~r[l\dt](t) ~r[l\dt](t-l\dt)
			\label{eq:GARTCH}
\end{eqnarray} 
with the 4 parameters $\sigma, \wInfty, \mu, \theta$, 
and the integer lag parameters $l$.
For $\theta = 0$, these equations reduce to the usual GARCH(1,1) process.
The rational for the equations is the following.
The ``internal'' state $\sigma_1$ measures the volatility at the time horizon $\tau = -\dt/\ln(\mu)$,
which is the volatility computed or perceived by a group of market participants.
The equation \ref{eq:GARTCH} models their trading pattern which depends on 
the difference with the (expected) mean volatility, 
and on the recent trend/drift of the price.
The ``parameter'' $l$ can be chosen {\it a priori}
with the guidance of the empirical analysis above.
It can also be studied systematically as done in sec.\ref{subsec:GARTCHLag}.

\subsection{The I-GARTCH(1) process}
For $\wInfty = 0$, the GARTCH(1,1) equations reduce
to the linear I-GARTCH(1) process
\begin{eqnarray}
    \sigma_1^2(t) & = & \mu\;\sigma_1^2(t-\dt) + (1 - \mu)r^2(t)  \\
      \sigmaEff^2(t+\dt) & = & \sigma_1^2(t) + \theta ~r[l\dt](t) ~r[l\dt](t-l\dt)
\end{eqnarray}
with two parameters $\mu$ and $\theta$.
A slight variation consists in using the definition
\begin{equation}
      \sigmaEff^2(t+\dt)  = \mu\sigmaEff^2(t) + (1 - \mu)r^2(t) 
      	+ \theta ~r[l\dt](t) ~r[l\dt](t-l\dt)
\end{equation}
In practice, both definitions give very similar results.
The empirical results on the 1 day forecast accuracy 
have been computed with the second definition.

We have also included in the study the RiskMetric volatility.
This process corresponds to the I-GARCH process with a fixed parameter $\mu$.
As we are working with hourly data (in business time scale),
we take $\mu = 0.93^{1/24}$.

\subsection{The Long Memory volatility processes}
The long memory process has been introduced in \cite{GOZ.2003-10-24}.
It incorporates  in a minimal way the power law decay of the lagged correlation for the 
absolute value of the return, or of the square return. 
This model is structureless with respect to the time horizons, 
namely it has a uniform structure between a lower and upper cut-off.
Therefore, it does not include the specific  market components 
as observed and modeled in \cite{GOZ.2000-11-06,LZ.2003}.
In the empirical analysis below, we have used the ``microscopic'' 
version of the long memory process (see \cite{GOZ.2003-10-24}). 
The inclusion of a trend term in the long memory process should preserve 
this simple and uniform structure.
Therefore, the idea is to include one 
$r\cdot L[r]$ term for each partial volatility $\sigma_k$,
and to have weights given by a simple power law.

The long memory models are built with a set of (historical) volatilities
computed  over a set of time horizons increasing as a geometric series:
\begin{eqnarray}
    \tau_k            & = & l_k\tau_0  \hspace{4em} k = 1,\cdots, n  \nonumber\\
    \mu_k            & = & \mu_0^{1/l_k} = \exp\left(-\;\dt/\tau_k\right)   \\      \sigma_k^2(t) & = & \mu_k\sigma_k^2(t-\dt) + (1 - \mu_k)r^2(t). \nonumber
\end{eqnarray}
The time horizons $\tau_k$ correspond to the characteristic times
of the EMA at which the historical volatility is measured.
The time structure $l_k$ of the process is a geometric series $l_k = \rho^{k-1}$,
with the progression of the series chosen to be $\rho = 2$ in this work.
The base time scale $\tau_0$ corresponds to the shortest time scale at
which a volatility is measured, and is one of the process parameters.
The empirical studies in this work have been done with hourly data and with $n=12$ components,
corresponding to an upper cut-off of 6 months.
The effective volatility $\sigmaEff$ for the long memory (LM) affine (Aff) 
processes LM-Aff-ARTCH($n$),
with $n$ components and trends, is
\begin{eqnarray}
    \sigmaEff^2(t+\dt) & = & \sigma^2 + (1 - \wInfty) \sum_{k = 1}^n \chi_k \left( \sigma_k^2(t) - \sigma^2 \right)
    		+ \sum_{k = 1}^n \theta_k ~r[l_k](t)~r[l_k](t - l_k \dt)
    \label{eq:def-LM-Aff-ARCH} \nonumber\\
        & = & \sum_{k = 1}^n w_k \sigma_k^2(t) + \wInfty \sigma^2 
		+ \sum_{k = 1}^n \theta_k ~r[l_k](t)~r[l_k](t - l_k \dt) \\
    \chi_k & = & c ~\rho^{- k\lambda} = c\left(\frac{1}{l_k}\right)^{\lambda}
           \hspace{4em} \text{with}\hspace{1em} 1/c = \sum_{k=1}^n \rho^{- (k-1)\lambda} \nonumber\\
    w_k & = & (1 - \wInfty) \chi_k \nonumber \\
    \theta_k & = & \theta_0 ~\rho^{- (k-1)\lambda'} = \theta_0 \left(\frac{1}{l_k}\right)^{\lambda'}
\end{eqnarray}
The ``normalization constant'' $c$ is chosen so that $\sum\chi_k = 1$
and $\sum w_k + \wInfty = 1$.
The ``mean terms'' introduce two constants $\sigma$ and $\wInfty$. 
For $\wInfty = 0$, the linear model is obtained.
The ``trend terms'' depend on the two constants $\theta_0$ and $\lambda'$.

In the comparative study, the long memory models allow us
to compare the processes with short memory (exponential) and long memory (power law).
We include 4 versions of the long memory process, 
namely one linear model LM-Lin-AR(T)CH (linear, similar to I-GAR(T)CH, 
but with long memory)
and one affine model LM-Aff-AR(T)CH (affine, similar to GAR(T)CH, 
but with long memory).

\subsection{The Market Component volatility processes}
In \cite{GOZ.2000-11-06,LZ.2003}, the correlation between historical and 
realized volatility is computed for a range of time horizons, 
and a similar computation was done for the change in historical 
volatility versus the realized volatility.
These correlations show clearly the characteristic time horizons of the market participants,
essentially at intra-day, daily, weekly and monthly time intervals.
The observed heterogeneity of the volatility correlations 
can be reproduced with a process 
that incorporates the same characteristic time horizons.
As noted in the empirical section, the trend effect shows very similar 
characteristic time intervals.
In order to include the trend effect in the market components process, 
it is enough to add a trend term for each volatility component.
The market component ARCH process is presented in detail in \cite{LZ.2003},
we give here for completeness the definition.

For the market component model, instead of measuring the (historical) volatilities with a simple Exponential Moving Average (EMA),
which has an exponential kernel, we use an MA operator
which has a more rectangular-like kernel.
The MA operator is defined by \cite{GOZ.2001-11-01}:
\begin{eqnarray}
    \MA[\tau, m; z](t) & = & \frac{1}{m} \sum_{j=1}^m \EMA_j(t)        \label{eq:MA} \\
    \EMA_1(t)  & = & \mu\EMA_1(t-\dt) + (1 - \mu)z(t)        \nonumber\\
    \EMA_j(t)  & = & \mu\EMA_j(t-\dt) + (1 - \mu)\EMA_{j-1}(t)  \nonumber\\
    \mu        & = & \exp\left(-\;\dt(m+1)/\tau\right)                \nonumber
\end{eqnarray}
The coefficient $\mu$ is computed from the time horizon $\tau$,
so that the memory length of the MA operator is $\tau$.
The parameter $m$ control the shape of the kernel,
and we take $m=8$ which gives a fairly rectangular kernel.

The Mkt-Aff-ARTCH process equations, with the trend terms, are:
\begin{eqnarray}
    \sigma_k^2(t) & = & \MA[\tau_k, m; r^2](t) \\
    \sigmaEff^2(t+\dt) & = & \sigma^2 + 
    	(1 - \wInfty) \sum_{k = 1}^n \chi_k \left( \sigma_k^2(t) - \sigma^2 \right) 
	+ \sum_{k = 1}^n \theta_k ~r[l_k\dt](t) ~r[l_k\dt](t-l_k\dt)
\end{eqnarray}
with the constraint
\begin{equation}
    \sum_{k = 1}^n \chi_k = 1.
\end{equation}
The linear version of the process Mkt-Lin-ARTCH is 
obtained by taking $\wInfty = 0$.

\subsection{Other processes}
For completeness, we include in the study a few processes 
which do not have an obvious extension to incorporate the trend effect.
The first one is the ``permanent forecast''. 
The historical volatility $\sigma_\text{hist}^2[\dts]$,
with equal weight on a time interval $\dts$, is 
\begin{equation}
    \sigma_\text{hist}^2[\dts](t) 
    	= \frac{1}{n}\sum_{t-\dts+\dtr \leq t' \leq t} r^2[\dtr](t').
\end{equation}
As all the points are equally weighted from $t-\dts$ to $t$, 
the memory kernel corresponds to a rectangular moving average.
The ``permanent forecast'' uses $\sigma_\text{hist}$ 
for the volatility forecast over any time horizon.

The I-GARCH(2) process is a natural extension of
GARCH(1,1), where the mean volatility $\sigma$ is replaced 
by an exponential moving average:  
\begin{eqnarray}
    \sigma_1^2(t) & = & \mu_1\sigma_1^2(t-\dt) + (1 - \mu_1)r^2(t) \\
    \sigma_2^2(t) & = & \mu_2\sigma_2^2(t-\dt) + (1 - \mu_2)r^2(t)   \nonumber\\
    \sigmaEff^2(t+\dt) & = & w\sigma_2(t)^2 + (1 - w) \sigma_1(t)^2(t).  \nonumber
\end{eqnarray}
This process is linear (in $\sigma^2$ and $r^2$), with two components. 
We name it I-GARCH(2) as it is a natural extension of I-GARCH(1).

\section{Volatility forecast using the processes}
\label{section:volForecast}
\subsection{The data set}
For all the empirical results presented in this section, 
the data set is a regular time series for USD/CHF, sampled
hourly in business time. 
This series is obtained by aggregation by a factor 28 of the homogeneous 
time series used in the empirical analysis.
The resulting sampling time interval of 1h24m corresponds to 7/5 of one hour, 
and is such that in average, 120 points per week are taken.
Essentially, during the business week from Monday to Friday, 
it corresponds to one point per hour, 
and no sampling point is taken during the week-end. 
Therefore, we will use the (imprecise) word ``hourly'' data when referring to this data set.
The data set is computed from 1.1.1989 to 1.7.2000.
The year 1989 is used for the build-up of the processes 
(i.e. the data for 1989 are inserted in the volatility 
processes so that they build their internal states and forget the initial conditions, 
but the cost functions do not include these data).
The following 10.5 years of data are used for the various studies.

\subsection{Volatility forecast and its measure of quality}
For a quadratic process, the coresponding volatility 
forecast can be obtained by conditional averages.
Essentially, at time $t$ and with the information set $\Omega(t)$, 
the forecasted volatility  at time $t+\Dt$ is given by the conditional average
\begin{equation}
  \Fcst{j\dt; \sigmaEff^2}(t) = \condExpt{\sigmaEff^2(t+\Dt)} 
\end{equation}
with $\Dt = j\dt$.
In order to compare with the realized volatility, 
one must compute the mean forecasted volatility between $t$ and $t + \Dt$
\begin{equation}
    \AvgFcst{\Dt; \sigmaEff^2}(t) = \frac{1}{m}\sum_{j=1}^m \Fcst{j\dt; \sigmaEff^2}(t)
\end{equation}
with $\Dt = m\, \dt$.
The processes above are estimated by minimizing the root mean square error $\text{RMSE}[\Dt, \theta]$
between the forecasted and realized volatility
\begin{equation}
    \text{RMSE}^2[\Dt, \theta] =
      \sum_t \left( \sqrt{\AvgFcst{\Dt, \theta; \sigmaEff^2}(t)} - \sigmaReal[\Dt](t) \right)^2
    \label{def:ForecastError}
\end{equation}
where $\theta$ is the set of process parameters.
One measure of quality used to compare the processes below 
is the relative RMSE, given by
\begin{equation}
	\text{rel.RMSE} = 1 - \frac{\text{RMSE}[\Dt]}{\text{stdDev}[\sigmaReal]}
\end{equation}
with $\text{stdDev}[x]$ the standard deviation of the time series $x$.
The relative RMSE measures the RMSE compared to the variance of the realized volatility, 
with values around 0 when the forecast is uncorrelated with the realized volatility, 
and value 1 for a perfect forecast
(i.e. it has a similar range as the linear correlation).
The other measure of quality used to compare processes 
is the usual linear correlation 
between the forecasted and realized volatility. 

\subsection{Study of the lag term in the I-GARTCH and GARTCH processes}
\label{subsec:GARTCHLag}
\begin{figure}[htb]
  \centering
  \includegraphics[width=0.8\textwidth]{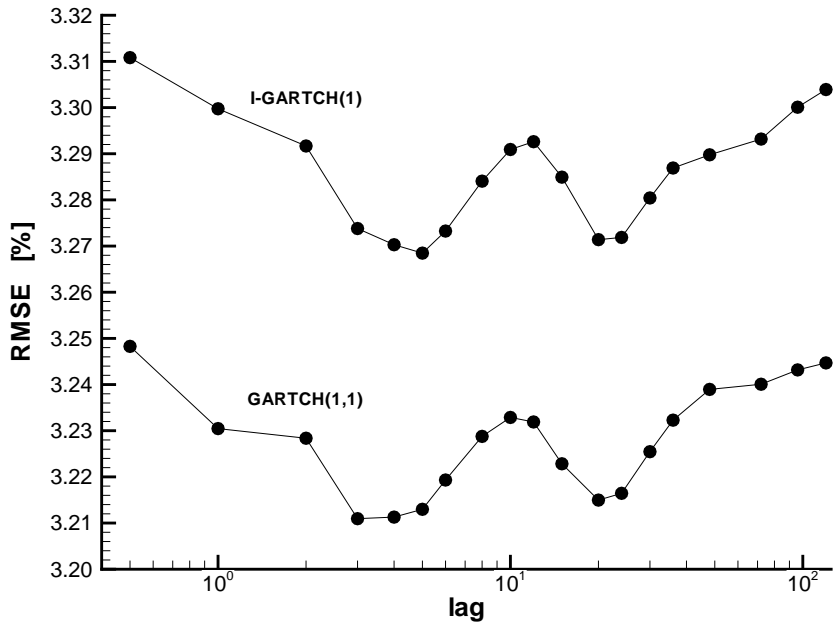}
  \caption{\small \sf The RMSE between the 1 day realized volatility and 
  the forecast for the daily volatility according to the I-GARTCH(1) and GARTCH(1,1) processes.
  The horizontal axis give the lag size $l$, 
  while the corresponding time interval is $l\dt$ with $\dt = $ 1 business hour.
  The 2 points at the extreme left of the curves, 
  located at the arbitrary location $l = 0.5$, 
  correspond to the usual I-GARCH(1) and GARCH(1,1) processes.
  The parameters of both processes are estimated on the data set, 
  for the different value of $l$, 
  and the RMSE at the minima is reported on the vertical axis. 
  }
  \label{fig:rLrLagStudy_GARTCH11}
\end{figure}

In the trend term $r[l\dt](t) ~r[l\dt](t-l\dt)$, the size of the lag $l\dt$ 
is effectively a parameter that must be fixed.
A simple solution is to use the empirical analysis and fix $l$ {\it a priori}
according to the maxima of the correlation values.
Another solution for the GARTCH and I-GARTCH model is to do a systematic 
estimation of the processes for various values of $l$, and to compare the results.
This is the aim of fig.~\ref{fig:rLrLagStudy_GARTCH11}.
Both curves show very clearly 2 minima, located around 3 to 5 hours, and 24 hours.
This is precisely in agreement with the maxima for the correlation 
given in sec.~\ref{sec:empiricalAnalysis}.

The 2 points located at the left end of both curves 
(with an abscissa corresponding to $l = 0.5$)
correspond to the usual I-GARCH(1) and GARCH(1,1) processes.
This figure also shows the improvement provided by the trend term, 
compared to the mean reversion, 
as mean reversion is included in GARCH but not in I-GARCH.
Roughly, the improvement provided by the 
trend term is of the order of 2/3 of the mean reversion,
namely it is slightly smaller that the mean reversion.
This shows that, in sample, 
the trend effect and mean reversion are of comparable magnitude.
In the comparative study below, we have used the trend term at 1 day.
The results using an intra-day trend are very similar.

\subsection{Comparison between processes for the volatility forecast error}
\label{subsec:fcstError}
Our goal in this section is to compare systematically the various processes above
in term of their forecasting performances, 
in order to measure the importance of the different ingredients included into the equations.
We proceed in two stages: first an in-sample comparison, 
which allows for a direct measure of the performances of the processes. 
In a second step, we measure the out-of-sample forecast performance with a continuous optimization, 
namely the parameters are estimated over a 5 years moving sample, 
and the forecast is computed out-of-sample just after the end of the estimation data.
This corresponds to the best setting that can be used in practice, 
with a continuous estimation of the parameters on the most recent 5 years of data.
This procedure measures the adequation of a process to compute a volatility forecast 
(as in the in-sample procedure), and at the same time the sensitivity of the parameters
with respect to the choice of the sub-sample.   
While the full in-sample procedure is simpler to interpret, 
the second one corresponds to the best setting that can be used in practice.

Taken together, both procedures can be used to build a measure
of the parameters robustness with respect to the choice of the data sample. 
We define the measure of robustness with respect to changing data sets by
\begin{equation}
    Q = \sum_t \left\| 
         \left(\AvgFcst{1d, 5y; \sigmaEff^2}(t)\right)^{1/2}
	 	- \left(\AvgFcst{1d, 10y; \sigmaEff^2}(t)\right)^{1/2}
                 \right\|
\end{equation}
where $\AvgFcst{1d, 5y; \sigmaEff^2}$ is a one day forecast for the realized variance,
with the parameters estimated on the subsample from $t - 5$y to $t$.
The term $\AvgFcst{1d, 10y; \sigmaEff^2}$ is the forecast 
with the parameters estimated on the full 10 years sample (in-sample forecast).
Essentially, the robustness factor $Q$ measures 
the difference in the forecasts when estimating the parameters
in a short sample (5 years) compared to a long sample (10 years),
and with the forecasts computed, respectively, out-of-sample and in-sample.
For large $Q$ values, the process is more dependent on the choice of the sub-samples, 
and therefore less robust.
The robustness factors are given in table~\ref{table:volFcst_robustness}.

\begin{table}
\begin{center}
\begin{tabular}{|l|D{.}{.}{-1}|D{.}{.}{-1}||D{.}{.}{-1}|D{.}{.}{-1}|r|}
\hline
 \multicolumn{3}{|c||}{no trend} & \multicolumn{3}{c|}{with trend} \\
\hline
\multicolumn{1}{|l}{process name} 
	& \multicolumn{1}{|c}{rel.RMSE} 
	&  \multicolumn{1}{|c||}{corr.} 
	& \multicolumn{1}{c}{rel.RMSE} 
	&  \multicolumn{1}{|c}{corr.} 
	& \multicolumn{1}{|r|}{process name} \\
\hline
\hline
mean in-sample volatility & -1.9 & 0.0  &      &      &  \\
\hline
permanent fcst(1d)        & -0.8 & 46.2 &      &      &  \\
permanent fcst(1w)        & 8.5  & 49.2 &      &      &  \\
permanent fcst(1m)        & 8.0  & 44.8 &      &      &  \\
\hline
RiskMetrics               & 10.5 & 47.7 &      &      &  \\
I-GARCH(1)                & 12.2 & 51.9 & 13.2 & 53.2 & I-GARTCH(1)  \\
GARCH(1,1)                & 13.8 & 52.6 & 14.7 & 53.8 & GARTCH(1,1)  \\
I-GARCH(2)                & 14.3 & 53.7 &      &      &  \\
\hline
LM-Lin-ARCH(12)       & 14.7 & 54.0 & 15.9 & 55.8 & LM-Lin-ARTCH(12) \\
LM-Aff-ARCH(12)       & 14.7 & 54.0 & 16.0 & 55.8 & LM-Aff-ARTCH(12) \\
\hline
Mkt-Lin-ARCH(4)       & 13.8 & 53.8 & 14.6 & 54.9 & Mkt-Lin-ARTCH(4)  \\
Mkt-Lin-ARCH(5)       & 14.4 & 53.9 & 15.3 & 55.3 & Mkt-Lin-ARTCH(5)  \\
Mkt-Lin-ARCH(6)       & 14.6 & 54.0 & 15.5 & 55.3 & Mkt-Lin-ARTCH(6)  \\
Mkt-Aff-ARCH(4)       & 14.7 & 54.0 & 15.5 & 55.2 & Mkt-Aff-ARTCH(4)  \\
Mkt-Aff-ARCH(5)       & 14.8 & 54.1 & 15.7 & 55.4 & Mkt-Aff-ARTCH(5)  \\
Mkt-Aff-ARCH(6)       & 14.8 & 54.1 & 15.7 & 55.4 & Mkt-Aff-ARTCH(6)  \\
\hline 
\end{tabular}
\caption{\label{table:volFcst_inSample}\sf
One day volatility forecast for the various processes.
The measures of quality of the forecast is the relative RMSE, in \%, 
and the linear correlation, in \%.
The parameters are estimated in-sample, on the full 11 years sample. 
The Mkt-*-AR(T)CH processes have 4, 5 or 6 components, with time horizons
of  4h48, 1 day, 1 week, 1 month; 
the 5 and 6 components models include a 3 months component; 
the 6 component model includes a 1 year component.
All multi-components models use a microscopic definition for the volatility
(see \cite{GOZ.2003-10-24} for a comparison with 
aggregated definitions of volatility).   
}
\end{center}
\end{table}

\begin{table}
\begin{center}
\begin{tabular}{|l|D{.}{.}{-1}|D{.}{.}{-1}||D{.}{.}{-1}|D{.}{.}{-1}|r|}
\hline
 \multicolumn{3}{|c||}{no trend} & \multicolumn{3}{c|}{with trend} \\
\hline
\multicolumn{1}{|l}{process name} 
	& \multicolumn{1}{|c}{rel.RMSE} 
	&  \multicolumn{1}{|c||}{corr.} 
	& \multicolumn{1}{c}{rel.RMSE} 
	&  \multicolumn{1}{|c}{corr.} 
	& \multicolumn{1}{|r|}{process name} \\
\hline
\hline
permanent fcst(5y)        & -5.3 & -2.0 &      &      &  \\
\hline
permanent fcst(1d)        &  0.3 & 46.9 &      &      &  \\
permanent fcst(1w)        &  7.2 & 47.3 &      &      &  \\
permanent fcst(1m)        &  5.5 & 40.6 &      &      &  \\
\hline
RiskMetrics               &  8.9 & 44.7 &      &      &  \\
I-GARCH(1)                & 10.7 & 49.9 & 11.9 & 51.1 & I-GARTCH(1)  \\             
GARCH(1,1)                & 12.6 & 50.8 & 13.4 & 51.9 & GARTCH(1,1)  \\          
I-GARCH(2)                & 13.5 & 52.4 &      &      &  \\
\hline
LM-Lin-ARCH(12)       & 14.2 & 53.0 & 15.7 & 55.0 & LM-Lin-ARTCH(12) \\
LM-Aff-ARCH(12)       & 14.0 & 52.8 & 15.5 & 54.9 & LM-Aff-ARTCH(12) \\
\hline
Mkt-Lin-ARCH(4)       & 13.1 & 52.3 & 14.1 & 53.6 & Mkt-Lin-ARTCH(4)  \\
Mkt-Lin-ARCH(5)       & 13.8 & 52.7 & 14.8 & 54.2 & Mkt-Lin-ARTCH(5)  \\ 
Mkt-Lin-ARCH(6)       & 14.0 & 52.6 & 14.9 & 54.1 & Mkt-Lin-ARTCH(6)  \\ 
Mkt-Aff-ARCH(4)       & 13.7 & 52.4 & 14.8 & 53.9 & Mkt-Aff-ARTCH(4)  \\ 
Mkt-Aff-ARCH(5)       & 13.8 & 52.7 & 14.8 & 54.2 & Mkt-Aff-ARTCH(5)  \\
Mkt-Aff-ARCH(6)       & 13.8 & 52.7 & 14.7 & 54.1 & Mkt-Aff-ARTCH(6)  \\
\hline 
\end{tabular}
\caption{\label{table:volFcst_outOfSample}\sf
As for table \ref{table:volFcst_inSample}, but with
the parameters estimated on a 5 years moving window, 
and the volatility forecast computed out-of-sample. 
}
\end{center}
\end{table}

\begin{table}
\begin{center}
\begin{tabular}{|l|D{.}{.}{-1}||D{.}{.}{-1}|r|}
\hline
 \multicolumn{2}{|c||}{no trend} & \multicolumn{2}{c|}{with trend} \\
\hline
\multicolumn{1}{|l}{process name} 
	&  \multicolumn{1}{|c||}{$Q$} 
	& \multicolumn{1}{c}{$Q$} 
	& \multicolumn{1}{|r|}{process name} \\
\hline
\hline
permanent fcst(5y)    & 0.77 &      &  \\
\hline
permanent fcst(1d)    & 0.0  &      &  \\
permanent fcst(1w)    & 0.0  &      &  \\
permanent fcst(1m)    & 0.0  &      &  \\
\hline
RiskMetrics           & 0.0  &      &  \\
I-GARCH(1)            & 0.11 & 0.09 & I-GARTCH(1)  \\             
GARCH(1,1)            & 0.28 & 0.28 & GARTCH(1,1)  \\          
I-GARCH(2)            & 0.15 &      &  \\
\hline
LM-Lin-ARCH(12)       & 0.10 & 0.10 & LM-Lin-ARTCH(12) \\
LM-Aff-ARCH(12)       & 0.14 & 0.13 & LM-Aff-ARTCH(12) \\
\hline
Mkt-Lin-ARCH(4)       & 0.07 & 0.08 & Mkt-Lin-ARTCH(4)  \\
Mkt-Lin-ARCH(5)       & 0.10 & 0.13 & Mkt-Lin-ARTCH(5)  \\ 
Mkt-Lin-ARCH(6)       & 0.12 & 0.19 & Mkt-Lin-ARTCH(6)  \\ 
Mkt-Aff-ARCH(4)       & 0.20 & 0.20 & Mkt-Aff-ARTCH(4)  \\ 
Mkt-Aff-ARCH(5)       & 0.17 & 0.17 & Mkt-Aff-ARTCH(5)  \\
Mkt-Aff-ARCH(6)       & 0.16 & 0.20 & Mkt-Aff-ARTCH(6)  \\
\hline 
\end{tabular}
\caption{\label{table:volFcst_robustness}\sf
The robustness factor $Q$, in \%, for processes without and with trend. 
Larger values indicate a stronger dependency on the sub-sample 
(i.e. less robust).
}
\end{center}
\end{table}

The results for the in-sample and for the continuous estimation 
with out-of-sample forecast are given respectively in table 
\ref{table:volFcst_inSample} and \ref{table:volFcst_outOfSample}.
Overall, the two setups produce consistent results, 
with the same ranking between processes.
From both tables, the major results about the structure of the processes are as follows.
\begin{itemize}
\item
The forecast accuracy improves as the memory kernel of the process approximates better the observed
power law decay of the lagged correlation. 
This leads to the improving sequence: ``permanent forecast'' (rectangular memory),
RiskMetric and I-GARCH(1) (exponential kernel), 
I-GARCH(2) (longer memory, but still exponential),
``long memory'' processes (power law kernel).
The shape of the memory kernel appears as the major classifying factor
in the table.

\item
The mean term does not improve much the forecast accuracy and is fragile.
This is a subtle point, which concerns the distinction between linear and affine processes 
(and between I-GARCH(1), GARCH(1,1) and I-GARCH(2)).
A mean term in the process equation for the effective volatility introduces two constants,
namely the mean volatility $\sigma$ and the corresponding coupling constant $\wInfty$.
These constants set the mean volatility as measured at an infinite time horizon,
and (likely) lead to well defined asymptotic distributions 
(see \cite{Nelson.1990} for a proof for GARCH(1,1) and 
\cite{GOZ.2003-10-24} for simulations with a long memory process).
By contrast, for the linear I-GARCH(1) process, 
the mean volatility is set by the initial conditions, 
and the asymptotic distribution is singular with all the mass at zero volatility. 
This singular asymptotic property is likely to 
be true for all linear processes (see the same references), 
but with a time to approach the asymptotic regime 
given by the longest time horizon in the process.
For long term Monte Carlo simulations, 
to have a well defined asymptotic distribution is essential,
and therefore affine processes must be used.
This property is different from the (power law) 
return to the mean after a large volatility spike,
and the inclusion of the ``return to the mean'' in a forecast.
For a process with multiple time horizons, 
the long term components act as a mean for the short time horizons. 
For example, forecast with an I-GARCH(2) process has a reversion toward 
the mean for time horizons between the 2 characteristic times of the EMAs. 
The comparison of the forecast qualities between the I-GARCH(1) process (no ``mean reversion''), and the 
GARCH(1,1) and I-GARCH(2) processes (both with ``mean reversion'' at the 1 day time horizon) 
shows that this ``reversion toward the longer term'' is quantitatively important.
Yet, I-GARCH(2) is better than GARCH(1,1), indicating 
that the constant mean volatility term is not the relevant factor.
Similarly, the results for the long memory processes show 
no differences between linear and affine   
versions, as all include a similar ``mean reversion'' at the 1 day forecast horizon 
(the longest time horizon in the process is of 6 months).

Moreover, the mean volatility is a parameter that is difficult to estimate, 
in the sense that it strongly depends on the chosen estimation sub-sample.
This dependency can be seen in the measure of the robustness $Q$, 
which is systematically larger for all affine processes, 
roughly by a factor $\sim$2 when compared to the corresponding linear processes.
Another case where the sensitivity of the mean volatility parameter 
can be observed is for the linear and affine
long memory processes: their forecast qualities are very similar
(even better out-of-sample for the linear process). 

To summarize this point, the ``reversion toward the mean'' must be included in a good forecast,
and it is better done with volatility components at long time horizons.
This advocates the use of linear processes for volatility forecasts
(like I-GARCH(2) or LM-Lin-ARCH processes), 
and to avoid using affine processes (like GARCH(1,1) process).
For Monte Carlo simulations at time horizons up to the largest 
characteristic time horizon included in the process 
(like in scenario simulations, or in risk evaluation), 
linear processes can be used.  
The affine processes must be used only for long term Monte Carlo simulations,
where the well defined asymptotic distribution should be close to the empirical volatility distribution.

\item
The trend effect is quantitatively important.
For all the processes, the relative RMSE is in the range 10\% to 14\%;
the addition of trend terms improves it by 1\% to 1.5\%. 
This is comparable to the improvement provided by a 
power law memory over an exponential memory.
Moreover, the robustness factors $Q$ are very similar when including trend terms 
(sometimes even smaller despite the larger number of parameters).
These forecast quantitative measurements clearly show that the 
trend effect is part of the empirical data 
and that its inclusion in a process improves the forecast accuracy.

\item For every process with trends, the trend coefficients are positive.
This is in agreement with the explanation given in the introduction
in terms of the behaviour of the market participants. 
This is also in agreement with the positive correlation 
between $r\cdot L[r]$ and the realized volatility,
as analyzed in sec.~\ref{sec:empiricalAnalysis}.

For the long memory processes with trend LM-*-ARTCH, 
the trend magnitude is $\theta_0 \simeq 0.18$, 
with an exponent $\lambda' \simeq 1.3$.

\item
The market models have performances that are similar to the long memory processes.
For the 4 components model, the number of components and related time horizons 
are taken as in \cite{LZ.2003}. 
The models with 5 and 6 components include further time horizons of 3 months and 1 year.
For these market models, the forecasting performances improve continuously
with the increasing number of components.
This improvement shows
the importance of the long time horizons, even for a forecast 
at the comparatively very short horizon of 1 day. 
Yet, whether in-sample or out-of-sample, 
it is difficult to do systematically better than the simpler long memory processes.
It indicates that the accurate modeling of the market components 
as measured in \cite{LZ.2003} is not crucial (for volatility forecast), 
and that a good caricature of the historical-realized volatilities correlations 
is enough to capture the main time structure of the memory.
With our setup, simple is better. 

\item
The use of a coarse grained definition of volatility 
does not seem to improve the forecast accuracy, as studied 
in details in \cite{GOZ.2003-10-24}.

\end{itemize}

\section{Conclusions}
Our systematic investigation of the empirical correlations between the 
trend term $r\cdot L[r]$ and the realized volatility 
is in agreement with our intuitive explanation according to the market participant behaviors:
price trends induce position changes that will create volatility.
Indeed, only zero or positive correlations are found in empirical data. 
Moreover, the correlation is large at the typical time horizons of the market participants,
namely intra-day, 1 day, 1 week (weak correlation), and 1 month.
This decomposition of the price behaviour in term of the time horizons of
the market participants is fully in line 
with historical-realized volatility correlations \cite{GOZ.2000-11-06,LZ.2003}.   

The inclusion of trend terms in a process presents no technical difficulties,
and improves clearly the forecast accuracy.
For forecasting purpose, the key factor for a efficient process is to reproduce 
the power law decay of the memory,
as is done by the LM-Lin-ARCH process.
Then, the trend term seems to be the next important factor.
For example, the relative RMSE for the processes 
``permanent forecast'' (7.2\%), I-GARCH(1) (10.7\%),
I-GARCH(2) (13.5\%), LM-Lin-ARCH (14.2\%) and LM-Lin-ARTCH (15.7\%)
show  well the successive improvement provided by the 
memory shape converging to an exponential, 
and finally by the trend term.
Notice that, compared to the simpler I-GARCH process (1 parameter),
the last process (4 parameters) improves by a good margin the forecast accuracy. 
On the contrary, the inclusion of a mean term in the volatility 
process (affine equation) does not better much the forecast.
Yet, the ``reversion toward the mean'' at the time horizon considered 
(1 day in this work)
improves the forecasting performances, 
but this ``reversion'' is better achieved with a process 
with multiple time horizons.
For example, I-GARCH(2) is better than GARCH(1,1) (both with 3 parameters),
and LM-Lin-ARCH is even better (with only 2 parameters).

\begin{table}
\begin{center}
\begin{tabular}{|l|D{.}{.}{-1}|D{.}{.}{-1}|}
\hline
\multicolumn{1}{|l}{process name} 
	&  \multicolumn{1}{|c|}{rel.RMSE} 
	& \multicolumn{1}{c|}{corr.} \\
\hline
\hline
permanent fcst(1d)    & -17.0 & 37.5  \\
permanent fcst(1w)    &  4.1  & 38.2  \\
\hline
RiskMetrics           & 11.1  & 37.0  \\
I-GARCH(1)            & 12.7  & 40.9  \\             
GARCH(1,1)            & 16.8  & 43.0  \\          
I-GARCH(2)            & 17.3  & 43.6  \\
\hline
LM-Lin-ARCH(12)       & 18.0  & 44.2 \\
\hline 
\end{tabular}
\caption{\label{table:volFcst_LM-ARCH_data}\sf
As for table \ref{table:volFcst_inSample}, with in sample optimization, 
for simulated LM-Aff-ARCH data. 
}
\end{center}
\end{table}
At the very end, the quantitative improvements when moving 
to more sophisticated processes and forecasts seems not
as important as one could hope it to be.
Considering that 20 years separate the original GARCH(1,1) process 
(12.6\% relative forecast error) from the LM-Lin-ARTCH (15.7\%) process,
the 3\% improvement in the relative forecast error is fairly small.
One of the acquired knowledge in those 2 decades 
is the use of high frequency data.
This knowledge is already included in the present work, 
which use deseasonalised hourly data.
By using high frequency data, we acquire more informations 
on the recent short term volatility, and subsequently improve short term forecasts.
Beside that better use of the information present in the data, 
it seems difficult to improve substantially the forecast accuracy.

In order to measure how good are our different processes in forecasting volatility,
we use simulated data. 
We generate the equivalent of 40 years of hourly data, 
using a LM-Aff-ARCH process. 
The residues are drawn from a Student distribution with 4.5 degree of freedom,
in agreement with a log-likelihood estimate on the USD/CHF data. 
Then, we estimate a few processes on this time series, and the results for the full in-sample estimates are reported in table~\ref{table:volFcst_LM-ARCH_data}.   
In particular, the estimation of the LM-ARCH process on the synthetic
data has (essentially) no misspecification, which gives a kind 
of absolute bound on the achievable forecasting performance.
The conforting news is that the relative RMSE are similar to 
the one obtained with the USD/CHF time series. 
It indicates that the long memory process is good at extracting 
the available informations to build a forecast, 
but that the (very) fat tail of the price innovation destroys 
a large part of the forecastability.
In this sense, the long memory processes are probably pretty 
close to the best that can be achieved.
On the other hand, the correlations differ significantly: 
$\sim$50\% with the empirical data versus $\sim$40\% with the synthetic data.
For the time series themselves, the means and standard deviations of the 
volatility are respectively 11.1\% and 4.5\% for the USD/CHF data,
and 10.1\% and 3.3\% for the LM-ARCH data.
The large difference for the standard deviations, 
as well as the differences for the correlations of the forecasts,
indicate that the LM-ARCH process is not yet a very good model
for the empirical data.

To summarize, this paper presents a first investigation of the trend term, 
with an empirical analysis using foreign exchange data,
and with a modelisation perspective in an ARCH set-up.
It shows that the trend effect is quantitatively important.
Therefore, it would be interesting to extend systematically the present study 
to data from different asset classes.
Another interesting direction consists in also including the tick rate.
In the intuitive explanation of the trend effect developed here, 
trends trigger trades, that generate in turn volatility.
Such effects could be directly measured 
using data from exchange traded instruments,
where limit orders and market orders are available.

\section{Acknowledgments}
The author thanks Olsen \& Associates, Zurich, Switzerland, for providing the data set.

\bibliography{volatilityConditionalOnPriceTrend}
\bibliographystyle{apalike}

\end{document}